\documentclass[onecolumn,showkeys,eqsecnum,aps,superscriptaddress, 11pt, pre ]{revtex4}
\usepackage{epsfig}
\usepackage{dcolumn}
\usepackage{xspace}

\begin{document}
\title{Non-linear dielectric effect in the isotropic phase above the
isotropic-cholesteric phase transition}
\vspace*{0.5cm}
\author{Prabir K. Mukherjee$^1$\protect\footnote[3]
{Corresponding author: Tel: +913322411977, E-mail address: pkmuk1966@gmail.com},
 Sumanta Chakraborty$^1$ and Sylwester J. Rzoska$^{2,3}$ \\
\vspace{0.3cm}
{\it $^1$ Department of Physics, Presidency University, 86/1 College Street,
Kolkata - 700 073, India}\\
{\it $^2$ Institute of Physics, Silesian University, Uniwersytecka 4, 40-007
Katowice, Poland} \\
{\it $^3$ Institute of High Pressure Physics PAS, ul. Sokolowska 27/39, 00-143 
Warsaw, Poland}}
\date{\today}

\begin{abstract}
\centerline{Abstract}
\vspace*{0.5cm}
\noindent
Using the Landau-de Gennes theory, the temperature, pressure and frequency 
dependence of the non-linear effect in the isotropic phase above the 
isotropic-cholesteric phase transition is calculated. The influence of 
pressure on the isotropic-cholesteric phase transition is discussed by varying the coupling 
between the orientational 
order parameter and the macroscopic polarization of polar cholesterics.   
Comparing the results of the calculations with existing data, we finally 
conclude that the model provides a description of the isotropic-cholesteric 
transition that takes all experimentally known features of 
the unusual negative and positive pretransitional effect in the isotropic 
phase of the system into account in a qualitatively correct way. 
\end{abstract}

\maketitle

\newpage

\section{Introduction}

The cholesteric (Ch) mesophase is locally very similar to the nematic (N) phase
except that it is composed of optically active molecules. When optically active
material form a nematic phase, the preferred direction of the long-molecular
axis varies in a direction throughout the medium in a regular way and displays
a continuous twist along the optic axis leading to a helical structure. Thus
the cholesteric phase is a distorted nematic phase.

While the research on the isotropic-nematic (I-N) phase transition has been going on over the past
few decades, there is also increasing interest in the isotropic-cholesteric
(I-Ch) transition \cite{degennes,cheng,yang,coates,harada,kats,brazovskii1,brazovskii2,brazovskii3,longa,muralidhar,seidin}. De Gennes \cite{degennes}
studied the pretransitional effect in the isotropic phase of cholesterics. He 
pointed out that the pretransitional effect in the isotropic phase emerges due 
to the presence of chiral aggregates. 
The I-Ch transition is believed to be first order \cite{cheng,yang,coates,harada}. In the isotropic phase of cholesterics the chirality exists only over a 
short temperature range \cite{cheng}.  Brazovskii 
et al \cite{brazovskii1,brazovskii2} predict a first order transition to the spiral
phase, occurring in a region of substantial manifestation of critical anomalies 
due to the effect of critical fluctuations. Longa et al \cite{longa} studied 
the phase diagrams of cholesteric liquid crystal with a 
generalized Landau-de Gennes theory. Muralidhar et al \cite{muralidhar} measured the ultrasonic 
absorption both in the cholesteric and isotropic phases as a function of 
both temperature and frequency. They observed the anomalous behavior of the 
absorption in both the phases.
Seidin et al \cite{seidin} studied the mean field phase diagram of a bulk cholesteric liquid 
crystal subjected to an externally applied field. 

The quantitative characteristic of the I-Ch transition delivered a series of 
linear 
and nonlinear dielectric permittivity investigations in chiral 
isopentylcyanobiphenyl (5*CB) carried out by Rzoska and co-workers 
\cite{rzoska1,rzoska2,rzoska3}. They measured the pressure dependence of the 
non-linear dielectric effect ($NDE$) in the isotropic phase. It was found that 
$NDE$ decreased strongly towards negative values on pressuring. This unusual 
behavior is opposite to the I-N and isotropic-smectic-A (I-SmA) transitions \cite{rzoska4,mukherjee,rzoska5,rzoska6,rzoska7}. 
Further it was observed that for the lower measurement frequency the $NDE$ 
strongly increases in the direction of positive values on approaching the 
I-Ch transition. For higher measurement frequencies the negative-sign 
pretransitional anomaly in the isotropic phase occurs. The clear evidence 
for the pretransitional behavior of the $NDE$ in the isotropic phase of the I-Ch 
transition was observed. Thus the I-Ch phase transition is associated with a
pronounced
pretransitional $NDE$ since the aligning electric field $\mathbf E$ couples to
the critical fluctuations.
The pressure study on the $NDE$
suggests that the same critical-like behavior with exponent $\alpha=0.5$ occurs 
for the I-Ch transition similar to the I-N and I-SmA transitions.

In spite of these experimental and theoretical efforts, there remains still number of key 
questions concerning the properties of the I-Ch transition. In particular, 
the natural supercooling of the isotropic phase and the anomalies in the 
temperature and pressure dependence of the precritical scattering of light and 
the $NDE$ require deeper investigation. It is 
worthwhile here to point out that the Landau theories described above are able 
to explain some of the main experimental observations at ambient 
pressure. There is no such theoretical study on the pressure effect on the 
I-Ch transition. 

The purpose of the present paper is to study the pressure effect on the I-Ch 
transition. We calculate the pressure and frequency
dependence of the $NDE$ in the isotropic phase above the I-Ch
phase transition. The model 
gives a correct description of the qualitative nature of the positive and negative
pretransitional behavior of the $NDE$ in the isotropic phase of the I-Ch 
transition.

\section{Theory}

In this section we consider the Landau-de Gennes theory of cholesteric liquid 
crystals. The cholesteric order parameter proposed
by de Gennes \cite{degennes,degennes1} is a symmetric, traceless tensor described by
$Q_{ij}=\frac S2(3n_in_j-\delta_{ij})$. Here $\mathbf{n}=(n_x,n_y,n_z)$ and
$S$ is the magnitude of the nematic order parameter.
In the geometry which is considered here the
cholesteric director is assumed to be $n_i=e_x\cos k_0z+e_y\sin k_0z$. Here the
direction of the helix axis is in the $z$ direction. Thus the structure of the
cholesteric liquid crystal is periodic with a spatial period given by
$L=\pi/|k_0|$.
The unique feature of 5*CB is that its structure resembles one of the 
classical liquid crystalline compounds 5CB. The compound 5*CB composed 
of strongly polar molecules. The external electric field induces a macroscopic 
polarization ${\bf P_m}$. Then the free energy of the polar cholesteric phase is 
a function of the orientational order parameter $Q_{ij}$ and the polarization 
${\bf P_m}$. Taking into account the relatively small value of the induced 
polarization, one can expand the free energy in powers of  
$Q_{ij}$ and ${\bf P_m}$  
\begin{eqnarray}
F &=&F_0+\frac 12AQ_{ij}Q_{ij}-\frac 13BQ_{ij}Q_{jk}Q_{ki}  \nonumber \\
&&{}+\frac 14C_1(Q_{ij}Q_{ij})^2+\frac 14C_2Q_{ij}Q_{jk}Q_{kl}Q_{li} 
\nonumber \\
&&+\frac 12L_1\nabla _iQ_{jk}\nabla _iQ_{jk}
+\frac 12 L_2\nabla _iQ_{ik}\nabla _jQ_{jk}-L_3\varepsilon
_{ijk}Q_{il}\nabla _kQ_{jl} \nonumber \\
&&+\frac 1{2\chi _0}\mathbf{P_m}^2  
+G Q_{ij}P_miP_mj
- P_miE_i   \label{free1}
\end{eqnarray}
where $F_0$ is the free energy of the isotropic phase. $A=a(T-T^{*}(P))$ and 
$T^{*}(P)$ is the critical
temperature for a hypothetical second order transition. 
All other coefficients, as well as $a$ and $\alpha _0$ are assumed to
be temperature independent. $\chi_0$ is the polarizability of the isotropic 
phase. The parameter $G$ represents the anisotropy of the 
polarizability in the cholesteric phase. 
$\gamma$ may be positive or negative. The 
term $\sim P_kP_nQ_{kl}Q_{nl}$ is 
neglected for the simplicity. $L_1$, $L_2$ and $L_3$ are the 
orientational elastic
constants. $\varepsilon _{ijk}$ is the Levi-Cevita antisymmetric tensor of the
third rank. The term proportional to $L_3$ violates parity and is
responsible for the formation of a helical ground state. We assume $L_3>0$.
It must be stressed that in this
manuscript we consider only the uniaxial case
and also neglect the blue phases for simplicity.

In accordance with the experimental phase diagram (shown in Fig.1),
$T^*(P)$ can be expanded as
\begin{equation}
T^*(P)=T_0+uP
\label{temp1}
\end{equation}
where $u$ is a positive constant.

Now we consider the phases in which the nematic order is 
spatially homogeneous, $i.e.$ $S=$const. Equation (\ref{free1}) can be 
simplified if one assumes that the polarization is aligned along the nematic 
director $\bf n$ i.e. ${\bf P_m}=(0,0,P_m)$. We choose ${\bf E}=(0,0,E)$.
The substitution of $Q_{ij}$, and $\mathbf{P_m}$ in Eq. (\ref
{free1}) leads to the free energy 
\begin{eqnarray}
F &=&F_0+\frac 34AS^2-\frac 14BS^3+\frac {9}{16}CS^4
+\frac {1}{2\chi _0}P_m^2 -\frac 12G P_m^2S \nonumber \\
&&+\frac 94LS^2k_0^2-
\frac 94L_3S^2k_0
-P_mE
\label{free2}
\end{eqnarray}
where $C=(C_1+C_2/2)$ and $L_1=-2L_2/3\equiv L$.

After minimizing the free energy (\ref{free2}) with respect to $k_0$, the wave 
vector of the helix in
the cholesteric phase can expressed as 
\begin{equation}
k_0=\frac {L_3}{2L_1}
\label{wave1}
\end{equation}
Equation (\ref{wave1}) shows that $L_3>0$ is required in order to obtain a
positive value of $k_0$ in the cholesteric phase. However, the wave vector
$k_0$ of the cholesteric phase may be positive or negative, since there are two
possible ways
of rotation of the cholesteric helix.
Minimizing the free energy (\ref{free2}) with respect to the polarization $P_m$
leads to:
\begin{equation}
P_m=E\chi _0M  \label{pola1}
\end{equation}
where $M=(1-G \chi _0S)^{-1}$. 

The substitution of $k_0$ and $P_m$ from Eqs. (\ref{wave1})-(\ref{pola1}) into
Eq. (\ref{free2}), we obtain

\begin{equation} 
F=F_0^{*}+\frac 34
A^{*}S^2-\frac 14BS^3+\frac {9}{16}CS^4-\frac 12 \gamma \chi_0^2 E^2S-\frac 12 
G^2 \chi_0^3E^2-\frac{1}{2}G^3 \chi_0^4E^2S^3  \label{free3}
\end{equation}
The renormalized coefficients are

$F_0^* =F_0 - \frac{E^2\chi_0}{2}$,

$A^*=A-\frac {3L_3^2}{4L}.$

First we consider the I-Ch transition in the absence of external electric 
field.  The conditions for the first order I-Ch transition can
be obtained as
\begin{equation}
F(S)=0,F^{\prime}(S)=0, F^{\prime \prime}(S)\ge 0
\label{cond1}
\end{equation}
{\bf The temperature and pressure dependence of $S$ can be expressed as
\begin{equation}
(S-S^+)^2=S^{+2}-\frac {a}{3C}[T-up-T_0-(3L_3^2/4La)]
\label{order}
\end{equation}
Thus $S$ changes with the change of pressure in the cholesteric phase.}
When $S$ is fixed, Eq. (\ref{order}) may be rewritten as 
\begin{equation}
T(P)=T_{10}(S)+uP
\label{temp2}
\end{equation}
where
\begin{equation}
T_{10}(S)=T_0+(3L_3^2/2La)-(3C/2a)S(S-S^{+})
\label{temp2a}
\end{equation}
where $S^+=B/6C$.  
{\bf However a change on $S$, Eq. (\ref{order}) shows the nonlinearity of $T$ vs. $P$.}
The jump of the
orientational order parameter at the I-Ch transition is given by
\begin{equation}
S_{I-Ch}=\frac {2B}{9C}
\label{jump1}
\end{equation}
The I-Ch transition temperature is given by
\begin{equation}
T_{I-Ch}=T_0+\frac {B^2}{27aC}+\frac {3L_3^2}{4La}+uP
\label{temp3}
\end{equation}

The analysis of Eq. (\ref {free3}) shows the influence of the external 
electric field on 
the cholesteric liquid crystal results in two main effects. First, the electric 
field produces a shift of the transition temperature $T^*_{I-Ch}$ which is 
proportional to the square of the electric field   
\begin{equation}
\Delta T^*_{I-Ch}(P)=mE^2 \label{temp4}
\end{equation}
with $m=\frac {G^2\chi_0^2}{3a}$.
Secondly, the external electric field induces weak orientational ordering in the 
isotropic phase. The orientational order parameter induced by an electric
field in the
isotropic phase is calculated to a first approximation ($B=0$ and $C=0)$ 
and can be expressed as
\begin{equation}
S(E)=\frac {U}{(T-T_0-\frac {3L_3^2}{4La}-uP
)}E^2+\frac {V}{(T-T_0-
\frac {3L_3^2}{4La}-uP
)^2}E^4 \label{induced}
\end{equation}
where 

$U=\frac {G\chi_0^2}{3a}$,
 
$V=\frac {2G^3\chi_0^5}{9a^3}$.

Note that in the first approximation, $S(E)=\frac {U}{(T-T_0-\frac {3L_3^2}{4La}
-uP)}E^2$. 

We will now calculate the frequency dependent $NDE$ in the isotropic phase of 
the I-Ch transition. The equation of motion, which we will write down as  
torque balance equation, can now be written 
\begin{equation}
-\frac {\partial F}{\partial S}-\Gamma \frac {dS}{dt}=0
\label{balance}
\end{equation}
where $\Gamma$ is the rotational viscosity. Introducing a time dependence of the 
order parameter $S$ and the electric field $E$ as $S=S_0e^{j\omega t}$ and 
$E=E_0e^{j\omega t}$, Eq. (\ref{balance}) to a first approximation 
($B=0$ and $C=0)$ give
\begin{equation}
-\frac 32A^*S+\frac 12 G \chi_0^2E^2-j\omega \Gamma=0 
\label{balance1}
\end{equation}
Solving Eq. (\ref{balance1}), we get the frequency dependent induced 
order parameter in the isotropic phase as 
\begin{equation}
S(\omega)=\frac {U}{(T-T_0-\frac {3L_3^2}{4La}-uP+j\omega 
\frac {2\Gamma}{3})}E^2
\label{induced1}
\end{equation}

The $NDE$ denotes the change in the dielectric permittivity of a material that
originates  from the application of
strong static electric field $\mathbf E$. The $NDE$ is widely analogous to the
electro-optic Kerr effect which applies to the case of optical frequencies.
An anisotropy property is proportional to the induced order.
Hence the dielectric permittivity in the isotropic phase can be expressed as
\begin{equation}
\Delta \varepsilon (E)=\varepsilon (E)-\varepsilon (0)=\left( \Delta 
\varepsilon _f\right) _{m
ax}\,S(E,\omega).
\label{die1}
\end{equation}
where $\varepsilon (E)$ and $\varepsilon (0)$ are the dielectric
permittivities in a strong ($E$) and weak (measuring) electric field. 
$(\Delta \varepsilon _f)_{\max }$ denotes the maximum
anisotropy of the dielectric permittivity for the given frequency $f$.
Combining Eqs. (\ref{induced1}) and (\ref{die1}) we find
\begin{equation}
\varepsilon_{NDE}=\frac {\varepsilon (E)-\varepsilon (0)}{E^2}=
\varepsilon_{NDE}^{Re}-j\varepsilon_{NDE}^{Im}
\label{die2}
\end{equation}
where 
\begin{equation}
\varepsilon_{NDE}^{Re}=\frac {W}{(T-T_0^*)+\frac 
{4\omega^2
\Gamma^2}{9a^2(T-T_0^*)}}
\label{redie2}
\end{equation}
\begin{equation}
\varepsilon _{NDE}^{Im}=
\frac {2\omega \Gamma G\chi_0^2}
{9a^2[(T-T_0^*)^2+\frac {4\omega^2
\Gamma^2}{9a^2}]}
\label{imadie2}
\end{equation}
where $W=(\Delta \varepsilon _f)_{\max }U$ and $T_0^*=T_0+\frac {3L_3^2}{4La}+uP$.

{\bf To parameterize the pretransitional anomaly in 5*CB, Rzoska et al 
\cite{rzoska3} measured the apparent amplitude $A_{NDE}(T)$ of the $NDE$. From the relation 
(\ref{redie2}), the apparent amplitude can be expressed as 
\begin{equation}
A_{NDE}(T)=NDE\times (T-T_0^*)=W-\frac {4W\omega^2\Gamma^2}{9a^2(T-T_0^*)^2}
\label{amp}
\end{equation}
}
From an alternative point of view, we can take $A=a(P^*-P)$. 
$P^*$ is supercooling pressure. In this case $\varepsilon _{NDE}$
can be expressed as 
\begin{equation}
\varepsilon _{NDE}=\frac {\varepsilon (E)-\varepsilon (0)}{E^2}=
\varepsilon_{NDE}^{Re}-j\varepsilon_{NDE}^{Im}
\label{die3}
\end{equation}
where
\begin{equation}
\varepsilon_{NDE}^{Re}=\frac {W}{(P^*-\frac {3L_3^2}{4La}-P)+\frac {4\omega^2
\Gamma^2}{9a^2(P^*-\frac {3L_3^2}{4La}-P)}} \label{redie3}
\end{equation}
\begin{equation}
\varepsilon_{NDE}^{Im}=\frac {2\omega \Gamma G\chi_0^2}
{9a^2[(P^*-\frac {3L_3^2}{4La}-P)^2+\frac {4\omega^2
\Gamma^2}{9a^2}]}
\label{imdie3}
\end{equation} 

\section{Comparison with experiment}

In this section we will compare our theoretical results with the available 
experimental results. 
According to Eq. (\ref{temp3}), the I-Ch transition temperature $T_{I-Ch}$
increases with increasing pressure. The pressure dependence of the I-Ch
transition temperature of 5*CB was reported
by Rzoska et al \cite{rzoska2}. The experimental phase diagram for $T_{I-Ch}$ vs
$P$ is a straight line. From Eq. (\ref{temp2}), when $S$ is fixed, $T$ vs $P$
should be a straight line. Equation (\ref{temp3}) is fitted with our measured
data of $T_{I-Ch}$ as a function of pressure for 5*CB taking $(T_0+\frac {B^2}
{27aC}-\frac {3L_3^2}{4La})$ and $u$ as fit parameters. The fit (solid) line 
and the measured
data (solid circles) are shown in Fig. 1. The fit yields $(T_0+\frac {B^2}
{27aC}-\frac {3L_3^2}{4La})=250K$
and $u=0.364K/MPa$.
The temperature dependence of the real part of the $NDE$
for a fixed $P$ given by Eq. (\ref{die2}) can easily be verified with Fig. 6 of
Rzoska et al \cite{rzoska3}. As for the pressure and temperature dependence the
mean-field value of the critical exponent $\varepsilon _{NDE}^{Re} \propto
(P^*-\frac {3L_3^2}{4La}-P)^{-\gamma}\propto (T-T_0^*)^{-\gamma}$ gives 
$\gamma =1$ which agrees with
experimental observations. Furthermore, to check Eq. (\ref{redie2}), 
$\varepsilon _{NDE}^{-1}$ vs $T-T_0^*$ of Rzoska et al \cite{rzoska3} for 
constant pressure is plotted in Fig.2. The form of Eq. (\ref{redie2}) shows 
that there are 
several unknown parameters. It is unphysical to take all these parameters as 
fit parameters. We have, therefore, fitted Eq. (\ref{redie2}) with the 
measured $\varepsilon _{NDE}^{Re}(T)$ data using $W$, $\Gamma$  
and $\frac {4}{9a^2}$ as fit
parameters. The solid line in Fig. 2 is the best fit of the real part of 
Eq. (\ref{die2}) for the different frequencies. The values obtained for
the fit parameters are listed in Table I. {\bf The temperature dependence of the 
apparent amplitude of the $NDE$ in the isotropic phase 
for a fixed $P$ is shown in Fig. 3. The line is the fit to the 
theoretical expression Eq. (\ref{amp}) using $W$, $\frac {4}{9a^2}$ and 
$\Gamma$ as fit parameters. The values obtained for the fit parameters are 
$W=230.73\times 10^{-16}$ $m^2V^{-2}K$, $\frac {4}{9a^2}=0.92$ $K^2kHz^{-2}(MHz^{-2})
Poise^{-2}$ and $\Gamma=0.82$ $Poise$. 
The fit to the measured values are 
good in Fig. 1-Fig. 3. Table I shows that the rotational viscosity 
$\Gamma$ decreases with the increase of frequency as expected. This behavior 
clearly support the theoretical analysis. To the best of the authors knowledge 
there are still no experimental estimations of the frequency dependent 
rotational viscosity.}

\section{Conclusion}

We have examined the pressure effect on the I-Ch transition in the mean-field
description. The present analysis provides the first theoretical support to
the experimental observations of the $NDE$ in the isotropic phase of the I-Ch 
transition \cite{rzoska1,rzoska2,rzoska3}. The effect of 
pressure on the I-Ch transition is to increase 
the transition temperature with pressure.
The I-Ch transition is found to be a first order even at high
pressure. Our theory explain the unusual positive and negative pretransitional 
effect in the isotropic phase of the I-Ch transition. 
The same pretransitional phenomena is 
observed in the isotropic phase of the I-Ch transition similar to the 
I-N and I-SmA transitions. The critical 
exponent $\gamma
=1$ indicate the fluid like analogy in the isotropic phase of the I-Ch 
transition.

\section{Acknowledgments}

PKM thanks the Alexander von Humboldt Foundation
for equipment and book grant. SC
thanks the KVPY for Fellowship. SJR was supported by Ministry of Science and 
Education (Poland) Grant No. 202231737.

\newpage

\begin{table}[h]
\caption{Values of the observed frequencies and the corresponding
various fitted parameters in the isotropic phase of the I-Ch transition in
5*CB, as derived from a fit of Eq. (\protect\ref{redie2}) to the measured data
of Ref. [15].}
\label{nde}\bigskip
\par
\begin{tabular}{lcccl}
\hline
$\omega$ & $W$ & $\frac {4}{9a^2}$ & $\Gamma$\\
$$ &  $(10^{-16}m^2V^{-2}{K})$ & $(K^2kHz^{-2}(MHz^{-2})Poise^{-2})$ & ($Poise$)\\ \hline
$66 kHz$ & $233.99$ & $1.29$ & $1.13$\\
$3.2 MHz$ & $-123.86$ & $1.29$ & $0.96$\\
$6.6 MHz$ & $-151.84$ &$1.29$ & $0.86$\\  \hline
\end{tabular}
\end{table}

\newpage
\noindent
{\bf Figure Captions:}
\begin{figure}[h]
\caption{The pressure dependence of the I-Ch transition temperature.
The solid line is the best fit of Eq. (\protect\ref{temp3}).}
\label{fig1}
\end{figure}

\begin{figure}[h]
\caption{The temperature dependence of the $NDE$ in the isotropic
phase of 5*CB. The measured data are from Ref. \cite{rzoska3} 
and the line is the best fit of Eq. (\protect\ref{redie2}).}
\label{fig2}
\end{figure}

\begin{figure}[h]
\caption{The temperature dependence of the apparent amplitude $A_{NDE}$ of the 
$NDE$ in the isotropic
phase of 5*CB. The measured data are from Ref. \cite{rzoska3}
and the line is the best fit of Eq. (\protect\ref{amp}).}
\label{fig3}
\end{figure}

\newpage
FIG.1
\begin{figure}[h]
\centerline{\psfig{file=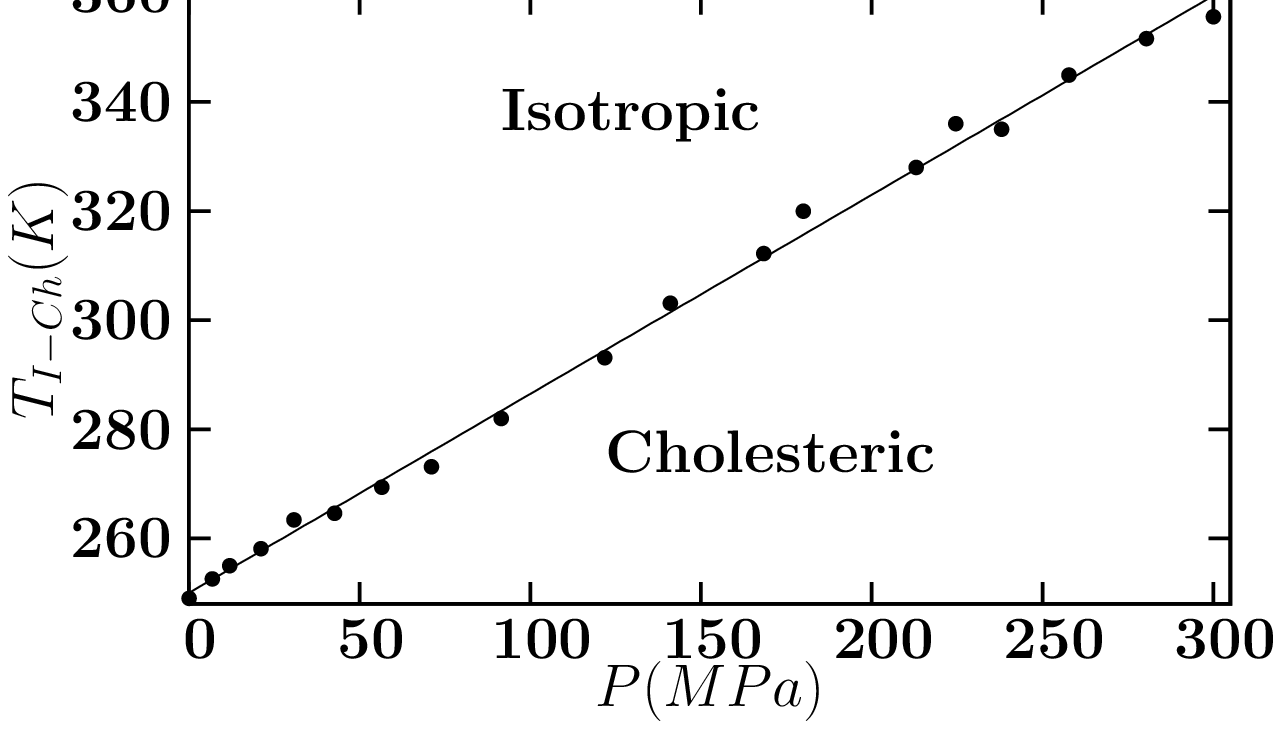,width=13cm, angle=0}}
\end{figure}

\newpage
FIG.2
\begin{figure}[h]
\centerline{\psfig{file=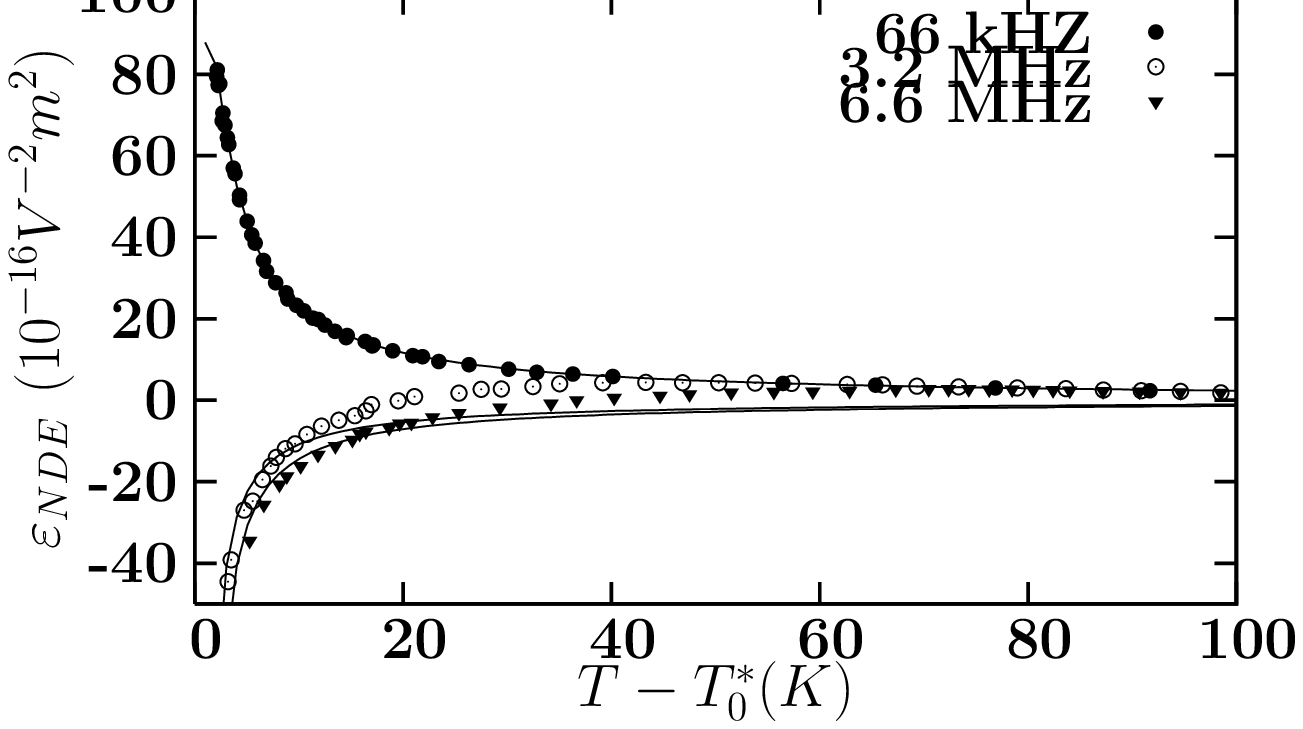,width=13cm, angle=0}}
\end{figure}

\newpage
FIG3.
\begin{figure}[h]
\centerline{\psfig{file=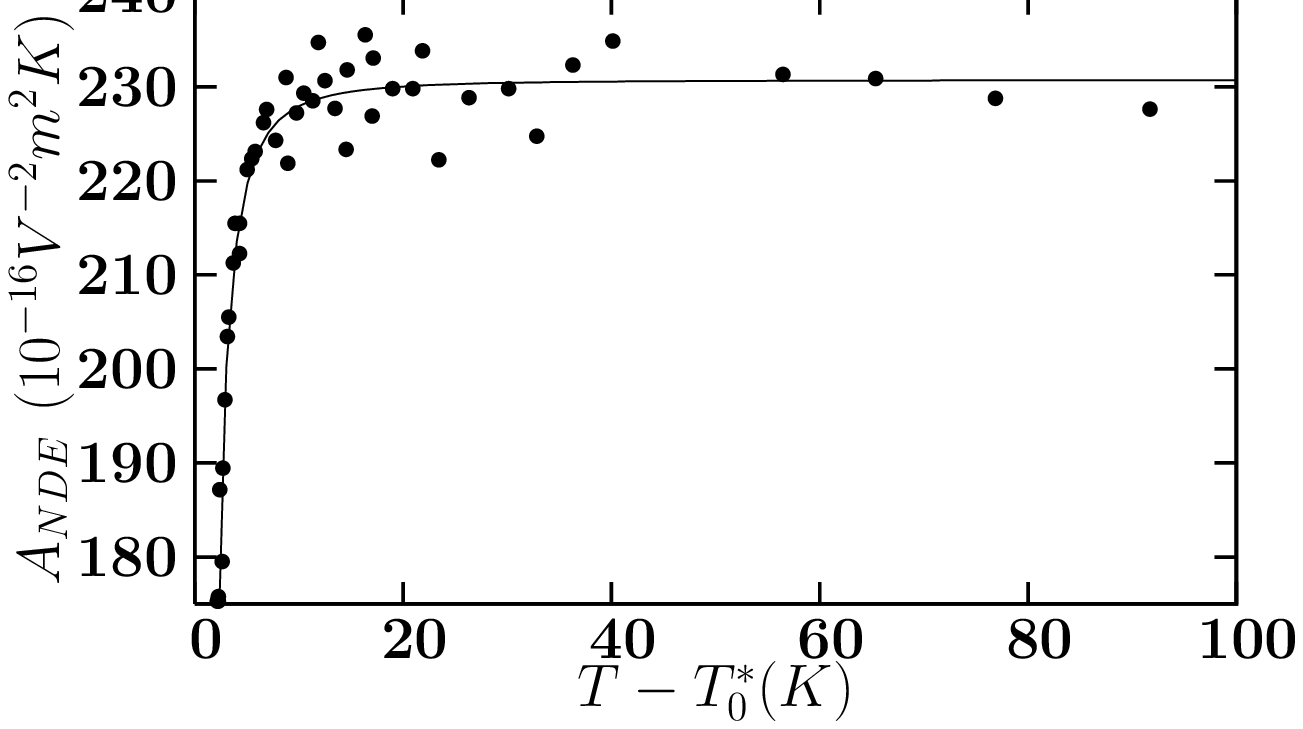,width=13cm, angle=0}}
\end{figure}

\end{document}